\documentclass[pra,twocolumn, superscriptaddress]{revtex4-2}
\usepackage{graphicx}
\usepackage{amssymb}
\usepackage{amsfonts}
\usepackage{amsmath}
\usepackage{lmodern}
\usepackage{mathrsfs}
\usepackage{mathdots}
\usepackage{collref}
\usepackage[colorlinks=true, citecolor=blue, urlcolor=blue]{hyperref}
\usepackage{tikz}
\usepackage{bm}
\usepackage{lipsum} 

\begin{document}

\title{Achieving Heisenberg scaling in low-temperature quantum thermometry}

\author{Ning Zhang}
\affiliation{Lanzhou Center for Theoretical Physics and Key Laboratory of Theoretical Physics of Gansu Province, Lanzhou University, Lanzhou 730000, China}

\author{Chong Chen}
\email{chongchenn@gmail.com}
\affiliation{Department of Physics, The Chinese University of Hong Kong, Shatin, New Territories, Hong Kong, China}

\begin{abstract}
We investigate correlation-enhanced low temperature quantum thermometry. Recent studies have revealed that bath-induced correlations can enhance the low-temperature estimation precision even starting from an uncorrelated state. However, a comprehensive understanding of this enhancement remains elusive. Using the Ramsey interferometry protocol, we illustrate that the estimation precision of $N$ thermometers sparsely coupled to a common low-temperature bath can achieve the Heisenberg scaling in the low-temperature regime with only a $\pi/2$ rotation of the measurement axis, in contrast to the standard Ramsey scheme. This result is based on the assumption that interthermometer correlations are induced exclusively by low-frequency noise in the common bath, a condition achievable in practical experimental scenarios. The underlying physical mechanism is clarified, revealing that the Heisenberg scaling arises from the intrinsic nature of the temperature, which is associated solely with the fluctuation of thermal noise. In contrast to the paradigm of independent thermometers, our proposed scheme demonstrates a significant enhancement in precision for low-temperature measurement, making it suitable for precisely measuring the temperature of ultracold systems.
\end{abstract}
\maketitle

\section{Introduction} 
Precise temperature control is crucial for cold-atom systems to ensure reliable quantum simulations \cite{Gross2017, Browaeys2020, Ebadi2021, Morgado2021}. Standard techniques for measuring temperature include time-of-flight \cite{Leanhardt2003,Gati2006} and impurity-based thermometry \cite{Spiegelhalder2009,Olf2015, Hohmann2016}. In contrast to the time-of-flight method, impurity-based thermometry offers a non-destructive alternative, attracting considerable attention for its potential advantages \cite{Mehboudi2019,Bouton2020,Mitchison2020,Adam2022}. However, in the low-temperature regime, impurity-based thermometry encounters the error divergence problem \cite{Mehboudi2019b, Potts2019, Jorgensen2020}.

 As illustrated in \cite{Zhang2023}, the temperature estimation precision is determined by both the heat exchange capacity and the correlation between the thermometer and the bath. Various quantum features that aim to increase heat exchange capacity or the correlation have been proposed to improve the precision of low-temperature measurement. These features include strong coupling \cite{Correa2017,Mehboudi2022, Mihailescu2023,Rodriguez2024}, quantum coherence \cite{Ullah2023}, quantum correlations \cite{Seah2019, Alves2022}, quantum criticality \cite{Hovhannisyan2018, Mirkhalaf2021, Aybar2022, Zhang2022}, and quantum non-Markovianity \cite{Zhang2021, Yuan2023, Xu2023}. Among them, bath-induced correlations between different thermometers have been shown to significantly improve the precision of low-temperature estimation by revealing new insights \cite{Gebbia2020, Planella2022, Brenes2023, Brattegard2024}. However, numerous abnormal characteristics have been elucidated in these schemes, e.g., these enhancements occur without initial correlations, are significant only at low temperature and weak coupling, and are ascertainable through local measurements. A comprehensive understanding of these characteristics remains an unsolved problem.

\begin{figure}[htb]
    \centering
    \includegraphics[width=1.0\columnwidth]{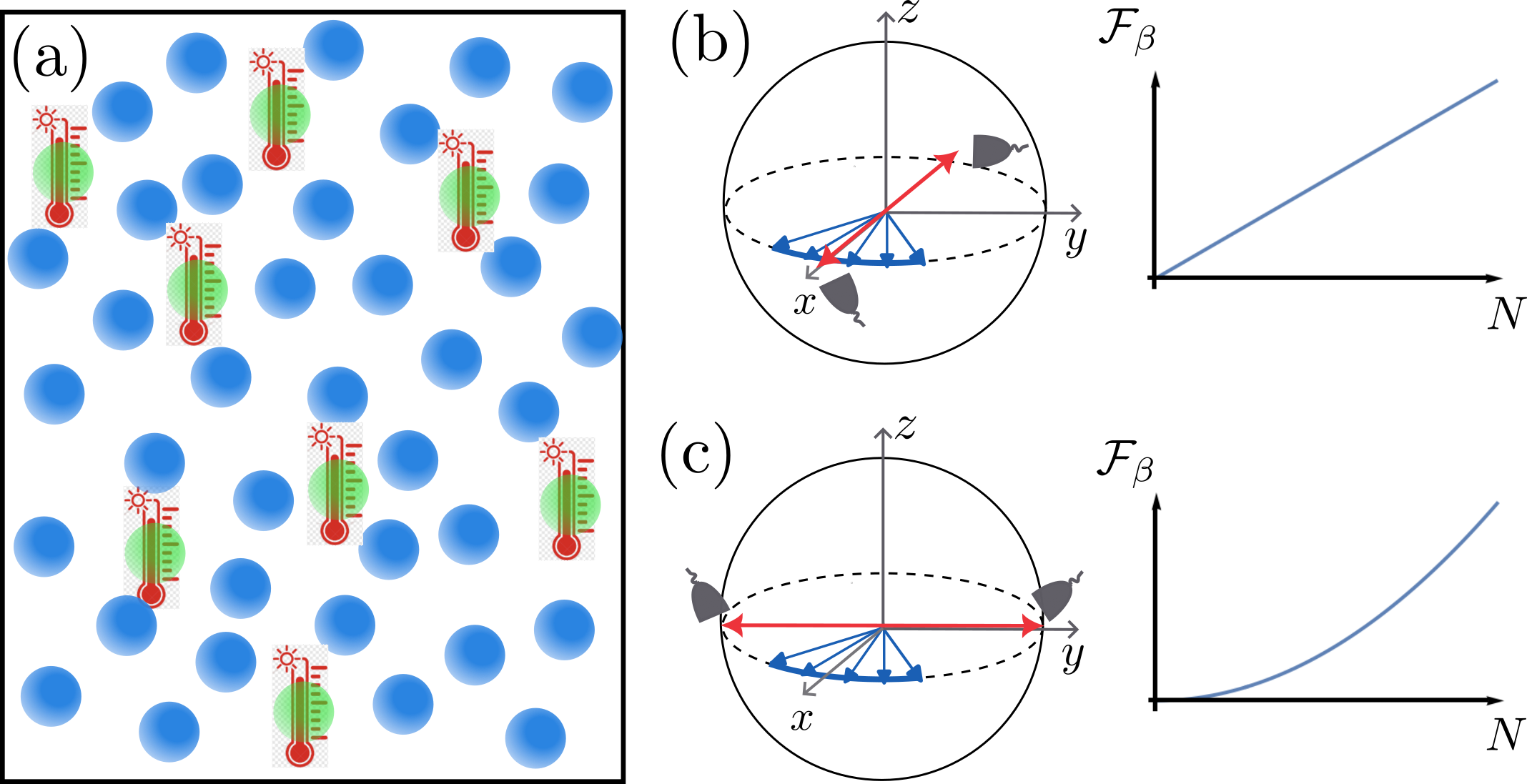}
    \caption{The scheme of correlation enhanced low-temperature quantum thermometry involves $N$ thermometers (green spheres) sparsely embedded into an low-temperature bath (blue spheres). Each thermometer is taken as a spin-$1/2$ and the temperature is measured based on the Ramsey interferometry protocol. The evolution state of the thermometer before measurement is depicted on the Bloch sphere (in the rotating frame). Our findings demonstrate that a linear scaling of the Fisher information with $N$ is achieved when the measurement axis is aligned along the $x$-direction (b), while an $N^2$ scaling is attainable when the measurement axis is oriented along the $y$-direction (c).}
    \label{fig:scheme}
\end{figure}

In this article, we address this problem by investigating $N$ thermometers sparsely embedded in a common low-temperature bath. Using the Ramsey interferometry protocol and under some simplifications, we provide an exact solution of this problem. We demonstrate that a Heisenberg scaling can be achieved in the low-temperature regime with only a $\pi/2$ rotation of the measurement axis compared to the standard Ramsey scheme. This result relies on the assumption that the correlations between the thermometers are induced exclusively by low-frequency noise within the bath--a condition that can be met by impurity-based thermometry in cold-atom systems. We show that the $\pi/2$ rotation changes the detecting method from independent measurement to correlation measurement, enabling Heisenberg scaling.  Furthermore, we elucidate that the $N^2$ scaling is intrinsically associated with the fundamental characteristics of temperature, namely that only fluctuations in thermal noise are temperature dependent. By clearly illustrating the mechanism of correlation-enhanced low-temperature thermometry, our work paves the way for accurately measuring the temperature of ultracold systems.

\section{$N$-thermometers in a common bath} Consider $N$ individual thermometers sparsely embed into a common ultracold bath with given temperature $T$, as shown in Fig. \ref{fig:scheme} (a). Such a configuration can be individual Cs atoms embedded into the ultracold bath of Rb. \cite{Hohmann2016, Schmidt2018,Bouton2020, Adam2022}.  Each of the thermometers is described by a spin-1/2 with the Hamiltonian $\hat{H}_S=\sum^{N}_{i=1}\frac{\hbar \omega_p}{2} \hat{\sigma}^{z}_{i}$, where $\hat{\sigma}^{z}_{i}$ denotes the Pauli matrices and $\omega_p$ represents the energy level splitting.  In the following, we would operate on the rotating frame and set $\hbar=1$. The bath may be any low-temperature system that interacts with the thermometers in the following manner:
\begin{equation}\label{eq:DephasingInteraction}
    \hat{H}_{I}(\tau)=\sum^{N}_{i=1}\hat{\sigma}^{z}_{i} \hat{X}_{i}(\tau),
\end{equation}
where $\hat{X}_{i}(\tau)$ represents the noise operator within the bath that acts upon the $i$-th thermometer. To facilitate our analysis, we postulate that the noise operator is bosonic and Gaussian in nature, which is an effective description of many quantum baths  \cite{Vega2017}, e.g., Bose gas \cite{Mehboudi2019,Planella2022,Khan2022} or Fermi Gas \cite{Mitchison2020, Brattegard2024}. In addition, the noise operator supports a decomposition based on its temperature dependence, expressed as $\hat{X}_{i}(\tau)=\hat{x}_{i}(\tau)+\hat{\zeta}_i(\tau)$, where $\hat{x}_{i}(\tau)$ and $\hat{\zeta}_i(\tau)$ represent the temperature dependent and independent parts, respectively. For the temperature dependent part, one can further induced a mode decomposition  as
\begin{equation}\label{eq:modeDecom}
    \hat{x}_{i}(\tau)=\sum_{k} g_{i,k} (\hat{b}_{k}e^{-i\omega_k \tau}+\hat{b}^{\dagger}_{k}e^{i\omega_k \tau}),
\end{equation}
where $\hat{b}^{\dagger}_{k}$ ($\hat{b}_{k}$) is the creation (annihilation) operator of $k$-th noise mode with characteristic frequency $\omega_k$. Each noise mode operates independently and is presumed to be in a thermal equilibrium state at temperature $T$, leading to $\langle \hat{b}^{\dagger}_{k}\hat{b}_{k'} \rangle_{B}=\bar{n}_{k}\delta_{k,k'}$, where $\bar{n}_{k}=\frac{1}{e^{\beta \omega_k}-1}$ with $\beta=\frac{1}{k_B T}$. 

The temperature estimation process follows the standard Ramsey interferometry protocol. By aligning $N$ thermometers along the $x$ direction, denoted as $\prod^{N}_{i=1}|+_{i,x}\rangle$, these thermometers undergo the evolution influenced by quantum noise $\{\hat{X}_{i}(\tau)\}$. After evolution time $t$, projection measurements $\hat{\Pi}_{s,\theta}$ along the direction $s \vec{e}_{\theta}=s(\cos \theta \vec{e}_{x}+\sin \theta \vec{e}_{y})$ with $s=\pm$ and $\theta \in [0,\pi/2]$ are applied to all thermometers, producing readouts $\{s_j\}$.  By repeating the Ramsey process $M$ times, the temperature is estimated from the probability distribution $P_{\bm{s}}$ of these readouts $\bm{s}\equiv \{s_j\}$. For local temperature estimation, the estimation precision is bounded by the Cram\'{e}r-Rao bound \cite{Paris2009}
\begin{equation}\label{eq:CRB}
    \Delta \beta^2 \ge 1/ M \mathcal{F}_{\beta},~~ \mathcal{F}_{\beta}=\sum_{\bm{s}}P_{\bm{s}} \mathcal{L}^2_{\beta},
\end{equation}
where $\mathcal{F}_{\beta}$ is the Fisher information and $\mathcal{L}_{\beta}=d \ln P_{\bm{s}}/d\beta$ is known as the score function. Given the clear relation between $\Delta \beta^2$ and $M$, we will omit $M$ for simplicity.

\section{Probability distribution} To calculate the probability distribution, we introduce the decomposition $|+_{x}\rangle \langle +_x|=\sum^{\pm}_{\eta,\bar{\eta}} |\eta_z\rangle \langle \bar{\eta}_z|$, where $|\pm_z\rangle$ denotes the states along the $\pm z$ directions. The evolution state of these thermometers before measurement is exactly solved, yielding
\begin{equation}\label{eq:rho}
    \rho_{\bm{\eta},\bar{\bm{\eta}}}(t)=\frac{1}{2^N}\exp[-\sum_{i,j} \Delta \eta_i \Gamma_{i,j}(t) \Delta \eta_j ],
\end{equation}
where $\bm{\eta} \equiv \{\eta_j\}$, $\bar{\bm{\eta}}\equiv \{\bar{\eta}_j\}$, $\Delta \eta_i \equiv \frac{1}{2}(\eta_i-\bar{\eta}_{i})$, and $\Gamma_{i,j}(t)=2\int^{t}_{0}d\tau_1\int^{t}_{0} d\tau_2 [\langle \{\hat{x}_{i}(\tau_1), \hat{x}_{j}(\tau_2)\}\rangle_{B}+\langle \{\hat{\zeta}_{i}(\tau_1), \hat{\zeta}_{j}(\tau_2)\}\rangle_{B}]$ denotes as the cooperative decay factor. Here $\{\hat{A},\hat{B}\}\equiv \frac{1}{2} (\hat{A}\hat{B}+\hat{B}\hat{A})$.   The measurements along the $\pm \vec{e}_\theta$ direction then yields (see Appendix \ref{app:A-I} for details)
 \begin{equation}
      P_{\bm{s}}=\langle e^{-\sum_{i,j} \Delta \eta_i\Gamma_{i,j}\Delta \eta_j +i\sum_{j} \Delta \eta_j (\theta+(1-s_j)\frac{\pi}{2})}\rangle_{\Delta \bm{\eta}},
 \end{equation}
 where $\langle \rangle_{\Delta \bm{\eta}}$ denotes the average over $\{\Delta \eta_{i}\}$ with probability $p_{\Delta \eta_j=\pm1} = \frac{1}{4}$ and $p_{\Delta \eta_j=0} = \frac{1}{2}$. By further utilizing $e^{-\sum_{i,j} \Delta \eta_i\Gamma_{i,j}\Delta \eta_j}\equiv \int^{\infty}_{-\infty} \frac{D \bm{\phi}}{Z} e^{- \sum_{i,j} \phi_i \Gamma^{-1}_{i,j} \phi_j +i 2\sum_{j} \phi_j \Delta \eta_j }$ with $Z=\sqrt{{\rm Det}[\pi \Gamma_{i,j}]}$ and $D\bm{\phi}=\prod_{i} d \phi_i$, $P_{\bm{s}}$ simplifies to
 \begin{equation}\label{eq:axulliaryField}
     P_{\bm{s}}=\langle  \prod_{j} \frac{1+s_j \cos(\theta+2\phi_i)}{2} \rangle_{\bm{\phi}},
 \end{equation}
 where $\langle \rangle_{\bm{\phi}}$ indicates the average over the auxiliary fields $\{\phi_j\}$ with the following distribution function $\frac{1}{\sqrt{\text{Det}[\pi \Gamma_{i,j}]}} e^{-\sum_{i,j} \phi_{i} \Gamma^{-1}_{i,j} \phi_{j}}$.  Physically, one can understand $2\phi_{i}$ as the rotation angle of the $i$-th thermometer under the quantum noise. Note that the rotation angle is a fluctuation quantity with a zero mean and a variance determined by the decay factor $\Gamma_{i,j}$.

Before proceeding with the calculation, it is crucial to clarify the characteristics of the decay factor $\Gamma_{i,j}$. Suppose that the $N$ thermometers are sparsely distributed within the low-temperature bath, as exemplified in the cold-atom experiments \cite{Schmidt2018, Bouton2020, Adam2022}. This sparse distribution implies that cooperative decay is predominantly associated with low-frequency noise modes, as contributions from different high-frequency noise modes tend to cancel each other out. Consider the temperature-dependent decoherence factor $\Gamma^{\beta}_{i,j}(t)\equiv 2\int^{t}_{0}d\tau_1\int^{t}_{0} d\tau_2 \langle \{\hat{x}_{i}(\tau_1), \hat{x}_{j}(\tau_2)\}\rangle_{B}$. Employing the mode decomposition presented in Eq. \eqref{eq:modeDecom}, we get
\begin{equation}\label{eq:Gamma-beta}
  \Gamma^{\beta}_{i,j}(t)  =4\int d\omega J_{i,j}(\omega) \coth \frac{\beta \omega}{2} \frac{1-\cos \omega t}{\omega^2},
\end{equation}
where $J_{i,j}(\omega)\equiv \sum_{k} g_{i,k} g_{j,k} \delta(\omega-\omega_k)$ represents the cooperative spectral density. A rough approximation of $J_{i,j}(\omega)$ is given by 
\begin{equation}
    J_{i,j}(\omega) \approx \left\{ \begin{array}{ll}
        J(\omega),  &  \text{ when } \omega \le \omega_{co}, \\
        J(\omega) \delta_{i,j},  & \text{ when } \omega > \omega_{co},
    \end{array} \right.
\end{equation}
where $\omega_{co}$ serves as the characteristic frequency that determines whether cooperative effects occur. Here, we assume that the low-frequency noise-induced correlations are long-range and that $J(\omega) \equiv \sum_{k} g^2_{k,i} \delta(\omega-\omega_k)$ is the same for all thermometers. Such approximations can be applied to the temperature-independent part, which yields 
\begin{equation}
\Gamma_{i,j}\approx \Gamma_{L}+\Gamma_{H} \delta_{i,j},
\end{equation}
where $\Gamma_{L}$ and $\Gamma_{H}$ represent the contributions from low-frequency and high-frequency parts, respectively.  Given this representation it can be established that the auxiliary fields exhibit mutual correlations as $\langle \phi_i \phi_j \rangle_{\bm{\phi}}=\frac{\Gamma_{i,j}}{2}$.  These correlations can be captured by a collective field $\varphi_0$ such that $\phi_i=\varphi_{i}+\varphi_0$ with $\langle \varphi^2_{0} \rangle=\frac{\Gamma_L}{2}$, $\langle \varphi_{i} \varphi_{j} \rangle=\delta_{i,j} \frac{\Gamma_H}{2}$ , and $\langle \varphi_{0} \varphi_{i}\rangle=0$. Under these considerations, the probability distribution finally reduces to
\begin{equation}\label{eq:probability}
     P_{\bm{s}}= \langle \prod_{j} \frac{1+s_j e^{-\Gamma_H}\cos(\theta+2\varphi_0)}{2}\rangle_{\varphi_0},
\end{equation}
where $\langle \rangle_{\varphi_0}$ denotes the average over $\varphi_0$ with the distribution function $\frac{1}{\sqrt{\pi\Gamma_L}}e^{-\frac{\varphi^2_0}{\Gamma_L}}$.  In the following, we would set $\Gamma_L \ll \Gamma_H$ due to the small contribution of the low-frequency noise and set $\Gamma_H \sim \mathcal{O}(1)$ for efficient temperature estimation.

Equation \eqref{eq:probability} reveals that the probability distribution is sensitive to the measurement angle $\theta \in [0,\pi/2]$. When $\theta=0$, a product probability distribution is obtained as (see Appendix \ref{app:A-II} for details)
\begin{equation}\label{eq:probability-I}
     P_{\bm{s}}= \prod^{N}_{j=1} [\frac{1+s_j e^{-\Gamma}}{2}+ \mathcal{O}(\Gamma^2_L)],
\end{equation}
where $\Gamma=\Gamma_L+\Gamma_H$. This is consistent with the findings derived from the independent thermometer scheme by setting $\Gamma_{i,j}=\Gamma \delta_{i,j}$. Temperature can be estimated from the mean value of $s_j$, reads $\langle s_j \rangle=e^{-\Gamma}$.  Given the absence of correlations among distinct thermometers, the temperature estimation can be efficiently executed via ensemble measurements, i.e, from the ensemble readout $S=\sum^{N}_{j=1} s_j$. On the other hand, when $\theta=\pi/2$, we have  (see Appendix \ref{app:A-II} for details)
\begin{equation}\label{eq:probability-II}
     P_{\bm{s}}= \frac{1}{z} [e^{S^2 \frac{\Gamma_L e^{-2\Gamma}}{1+2N\Gamma_L e^{-2\Gamma}}}+\mathcal{O}(N\Gamma^2_L)],
\end{equation}
where $z$ denotes the normalization factor. The property involving only the ensemble readout $S$ enables temperature estimation via ensemble measurements. For other cases with $0 < \theta < \pi/2$, the resultant probability distribution would be intermediate between the results shown in Eq. \eqref{eq:probability-I} and Eq. \eqref{eq:probability-II}.

In ensemble measurements, only the probability function $p(S)$ is relevant. By taking  $s_j$ as a binomial distribution, $p(S)$ is derived in the large $N$ limit, given by
\begin{equation}\label{eq:probability-S}
    p(S)\approx \left\{ 
    \begin{array}{cl}
        \frac{1}{\sqrt{2\pi N(1-e^{-2\Gamma})}}e^{-\frac{(S-N e^{-\Gamma})^2}{2N(1-e^{-2\Gamma})}}, &  \theta=0,\\
    \frac{1}{\sqrt{2\pi N(1+2e^{-2\Gamma}N\Gamma_L)}} e^{-\frac{S^2}{2N(1+2e^{-2\Gamma}N\Gamma_L)}}, & \theta=\frac{\pi}{2}.
    \end{array}
    \right.
\end{equation}
Using this result, one can explicitly derive the score function $\mathcal{L}_{\beta}$, calculate Fisher information $\mathcal{F}_{\beta}$, and then determine the fundamental limit of the estimation precision. 

\section{Precision analysis} Leveraging Eq. \eqref{eq:probability-S}, the derivation of the score function $\mathcal{L}_{\beta}$ is straightforward. Results are 
\begin{equation}\label{eq:ScoreF}
        \mathcal{L}_{\beta}\approx \left\{ 
    \begin{array}{cl}
        c_1(S-\langle S \rangle), &  \theta=0,\\
     c_2(S^2-\langle S^2\rangle), & \theta=\frac{\pi}{2},
    \end{array}
    \right.
\end{equation}
where $c_{1}=-\frac{e^{-\Gamma}}{1-e^{-2\Gamma}} \frac{d\Gamma}{d\beta}$ and $c_2=-\frac{1}{2N}\frac{d}{d\beta} (1+2e^{-2\Gamma}N\Gamma_L)^{-1}$. Note that the score function is usually associated with the observable employed in the parameter estimation \cite{Paris2009}.  We can conclude that the temperature for $\theta=0$ is inferred from the observable $S$, while for $\theta=\pi/2$, it is estimated from $\sum_{i<j}s_is_j$, given that $S^2=N+2\sum_{i<j}s_i s_j$.  In light of these results, we refer to the measurement scheme at $\theta=0$ as an independent measurement, while the scheme at $\theta=\pi/2$ is identified as a correlation measurement. This result elucidates that the observable employed in temperature estimation is measurement-axis dependent.

Similar to the cooperative effect, the accuracy of the temperature estimation is also influenced by the energy scale $\omega_{co}$. From Eq. \eqref{eq:Gamma-beta}, it can be observed that the temperature dependence of $p(S)$ originates from the function $\coth \frac{\beta \omega}{2}$ in $\Gamma^{\beta}_{i,j}$, which satisfies $\coth\frac{\beta \omega} {2}\approx (\frac{\beta\omega}{2})^{-1}$ for $\beta \omega \lesssim 1$ and $\coth \frac{\beta \omega}{2} \approx 1 $ for $\beta \omega \gtrsim 1$. Thus, for low-temperature estimation, where $\beta \omega_{co} \gtrsim 1$, only low-frequency noise is relevant. In contrast, at high temperature, both low-frequency and high-frequency noise are relevant. However, considering the degrees of freedom in the low-frequency regime are obviously fewer compared to those in the high-frequency regime,  the temperature estimation at high temperature is predominantly governed by high-frequency noise. Based on these analyses, for high temperature estimation, we have
\begin{equation}\label{eq:FI-HighT}
    \mathcal{F}_{\beta}\approx \left\{ 
    \begin{array}{cl}
        \frac{N}{e^{2 \Gamma}-1} |\frac{d \Gamma_H}{d\beta}|^2, & \theta=0,\\
        \frac{8(N\Gamma_L)^2}{(e^{2 \Gamma}+2N\Gamma_L)^2} |\frac{d \Gamma_H}{d\beta}|^2, & \theta=\frac{\pi}{2},
    \end{array}
    \right.
\end{equation}
where the contribution from $\Gamma_L$ is omitted.  The standard quantum limit, denoted $\mathcal{F}_{\beta} \sim N$, manifests itself when $\theta = 0$, attributed to the absence of interthermometer correlations. In contrast, a result $\mathcal{F}_{\beta}\le 2|\frac{d \Gamma_H}{d\beta}|^2$ is revealed in the case $\theta=\pi/2$. This indicates that correlation measurement deteriorates the estimation of temperature. For low temperature estimation, we have
\begin{equation}\label{eq:FI-LowT}
    \mathcal{F}_{\beta}\approx \left\{ 
    \begin{array}{cl}
        \frac{N}{e^{2 \Gamma}-1} |\frac{d \Gamma_L}{d\beta}|^2, & \theta=0,\\
        \frac{2 N^2}{(e^{2 \Gamma}+2N\Gamma_L)^2} |\frac{d \Gamma_L}{d\beta}|^2, & \theta=\frac{\pi}{2}.
    \end{array}
    \right.
\end{equation}
A Heisenberg scaling is reveled in the correlation measurement when $2N\Gamma_L \lesssim e^{2\Gamma}$. This condition can be satisfied by cold-atom systems, where a small number of thermometers are sparsely distributed within the ultracold bath. Thus, a Heisenberg scaling is expected to be observed in cold-atom systems. In addition, an $N$ independent result $\mathcal{F}_{\beta} \approx \frac{1}{2\Gamma_L^2}|\frac{d \Gamma_L}{d\beta}|^2$ is obtained in the large $N$ limit, indicating that there is no improvement when $2N\Gamma_L \gg  e^{2\Gamma}$.  In this case, one needs to go beyond the ensemble measurement by dividing these $N$ thermometers into $N/N_0$ groups with each $2N_0\Gamma_L \approx e^{2\Gamma}$. Applying correlation measurements for each group, we obtain the total Fisher information $\mathcal{F}_{\beta} \approx \frac{N N_0}{2e^{4 \Gamma}} \left| \frac{d \Gamma_L}{d\beta} \right|^2$. Compared to independent measurement, it shows a precision improvement of $(1-e^{-2\Gamma})e^{-2\Gamma}N_0/2\gg 1$ as $\Gamma_L \ll \Gamma_H$ and $\Gamma_H \sim \mathcal{O}(1)$. Hence, correlation measurement in the low-temperature regime can significantly enhance the  temperature estimation precision. Equations (\ref{eq:FI-HighT}, \ref{eq:FI-LowT}) represents the main conclusions presented in this study.

 From an information perspective, the Heisenberg scaling indicates that in the low-temperature regime, the primary thermal information sources from interthermometer correlations.  Consider two thermometers coupled to a common bath. The density matrix $\rho_{i,j}$ before measurement reads $\rho_{i,j}\approx \frac{1}{4}(I +e^{-\Gamma}\hat{\sigma}^{x}_i)(I +e^{-\Gamma}\hat{\sigma}^{x}_j)+\frac{1}{2} \Gamma_L e^{-2\Gamma}\hat{\sigma}^{y}_i\hat{\sigma}^{y}_j$. We know that only $\Gamma_L$ is temperature dependent at low temperature. Thus, both the single-thermometer term $e^{-\Gamma}\hat{\sigma}^{x}_i$ with $\Gamma=\Gamma_L+\Gamma_H$ and the inter-thermometer term $\frac{1}{2} \Gamma_L e^{-2\Gamma} \hat{\sigma}^{y}_i \hat{\sigma}^{y}_j$ are temperature dependent. However, because of the orthogonality between these two terms, one can detect either the first term through an independent measurement or the second term via a correlation measurement. For a system of $N$ thermometers, the independent measurement retrieves $N$ copies of thermal information, while $N(N-1)/2$ copies can be obtained through the correlation measurement. Henceforth, measuring these interthermometer terms exhibits superiority in low-temperature estimation. Physically, this superiority arises from the fundamental attributes of temperature, intrinsically linked to fluctuations in thermal noise $\langle\{\hat{x}_i (\tau_1), \hat{x}_j(\tau_2)\} \rangle$, rather than its average $\langle \hat{x}_i (\tau)\rangle$.  These noise fluctuations are precisely measured in the correlation measurement as $\langle s_i s_j\rangle \approx 2 \Gamma_L e^{-2\Gamma}$ and $\Gamma_L= 2\int^{t}_{0} d\tau_1 \int^{t}_{0} d\tau_2 [\langle\{\hat{x}_{L,i}(\tau_1), \hat{x}_{L,j}(\tau_2)\} \rangle+\langle\{\hat{\zeta}_{L,i}(\tau_1), \hat{\zeta}_{L,j}(\tau_2)\} \rangle]$, where $L$ indicates the low-frequency component. Thus, a significant improvement in accuracy is expected for the correlation measurement.  In contrast, for parameters that are independent of the noise fluctuations, e.g., the static magnetic field, correlation measurements would not bring any advantage.

\section{Numerical results} Consider the Ohmic noise spectrum $J(\omega)=\alpha \omega e^{-\omega/\omega_c}$ as a paradigmatic example to elucidate the results shown in Eqs. (\ref{eq:FI-HighT}, \ref{eq:FI-LowT}), where $\alpha$ characterizes the coupling strength and $\omega_c$ denotes the truncation frequency. According to the preceding analysis, we suppose $J_{i,j}(\omega)=J(\omega)$ when $\omega \le \omega_{co}$ and $J_{i,j} = \delta_{i,j}J(\omega)$ when $\omega > \omega_{co}$. Temperature-independent noise $\hat{\zeta}(\tau)$ is assumed to be white noise satisfying $\langle\{\hat{\zeta}(\tau_1), \hat{\zeta}(\tau_2)\} \rangle=\gamma \delta(\tau_1-\tau_2)$. In addition, the low-frequency contribution in $\hat{\zeta}(\tau)$ is ignored. The observable $O$ for estimating temperature is chosen in accordance with Eq. \eqref{eq:ScoreF}, adopting $S$ for the $\theta = 0$ case and $S^2$ for the $\theta=\pi/2$ case. The temperature precision is then determined from the signal-to-noise ratio \cite{Degen2017}, given by 
\begin{equation}\label{eq:SNR}
    \text{SNR}= \frac{|\delta \bar{O}|}{\Delta O},
\end{equation}
where $\bar{O}$ and $\Delta O$ denote the average and the fluctuation of $O$, respectively. By noting $\delta \bar{O}=\Delta \beta \frac{d}{\beta} \bar{O}$, it yields the ultimate estimation precision $\Delta \beta \ge \frac{\Delta O}{ |d\bar{O}/d\beta|}$, determined by the criterion $\text{SNR} \ge 1$. Note that $\frac{\Delta O^2}{ |d\bar{O}/d\beta|^2} \approx \mathcal{F}_{\beta}$.

\begin{figure}
    \centering
    \includegraphics[width=1.0\columnwidth]{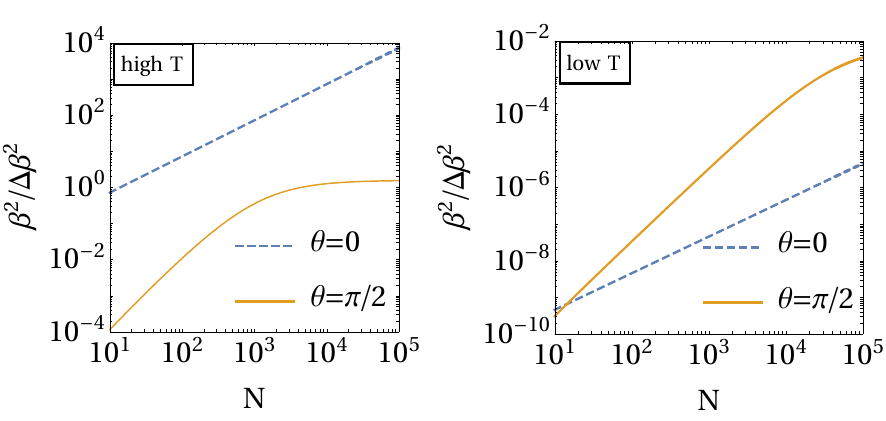}
    \caption{The temperature estimation precision $\beta^2/\Delta \beta^2$ for independent measurement, denoted as $\theta=0$, and correlation measurement, denoted as $\theta=\pi/2$, respectively. Parameters are: $\alpha=0.2$, $\omega_c=10.0$, $\omega_{co}/\omega_c=0.01$, $\gamma/\omega_c=0.1$, and $\beta\omega_c=10^3$ and $\beta\omega_c=1.0$ are selected for low-temperature and high-temperature cases, respectively. The evolution time is optimized for each case. In the high-temperature case, $t=0.1$ when $\theta=0$ and $t=0.15$ when $\theta=\pi/2$. In the low-temperature case, $t=0.6$ when $\theta=0$ and $t=0.18$ when $\theta=\pi/2$.}
    \label{fig:SNR}
\end{figure}

Figure \ref{fig:SNR} illustrates the precision of the temperature estimation in scenarios that encompass both high- and low-temperature under independent and correlation measurements. The results show that at high temperature, independent measurement, denoted by $\theta=0$, always achieves better performance. This observation aligns with our prior analysis, which stats that the precision of high-temperature estimation is predominanted by the high-frequency noise. However, interthermometer correlations are exclusively relevant for low-frequency noise.  In contrast, the correlation measurement shows an significant improvement at low temperature, revealing a Heisenberg scaling until the condition $2N \Gamma_L \lesssim e^{2\Gamma_L}$ is violated. Such an observation is consistent with the results shown in Eq. \eqref{eq:FI-LowT}, which state that the inter-thermometer correlation can be utilized to enhance the low-temeperature estimation precision.

\section{Discussion and conclusion}
In the maintext, we assume $\Gamma_L \ll \Gamma_H$. It is interesting to ask what happens when the cooperative decay occurs in both the low-frequency and high-frequency regimes such that $\Gamma_L > \Gamma_H$. A simple generalization Eq. \eqref{eq:FI-LowT} yields Fisher information in correlation measurement attains an $N$ independent value at low temperatures when $2N \Gamma_L \gg e^{2\Gamma}$. The correlation-enhanced signal $|\delta \bar{O}|$ in Eq. \eqref{eq:SNR} is counterbalanced by correlation-induced fluctuation $\Delta O$, resulting $N$ independent result. Thus, correlation measurements do not confer any benefit in this situation. Consequently, to realize precision enhancement, one needs the cooperative decay to be manifested exclusively in the low-frequency regime such that $\Gamma_L \ll \Gamma_H$. This condition can be satisfied when the thermometers are sparsely and randomly embedded within a high-dimension thermal bath, making the cooperative effect induced by high-frequency noise negligible. In addition, the infinite-long-range correlation is assumed within the main text. However, a realistic case is that the noise-induced correlation scales as $1/r^{\alpha}_{i,j}$ with $\alpha \ge 0$, where $r_{i,j}$ denotes the distance between $i$-th and $j$-th thermometers \cite{Defenu2023}. Assuming a uniform spatial distribution of thermometers within a thermal bath of $D$ dimensional, the relationship between Fisher information and $N$ diminishes to $\mathcal{F}_{\beta} \sim N  \rho_{0} \int  r^{-\alpha} dV \sim N^{2-\epsilon}$, where $\epsilon = \min(\alpha/D, 1)$ and $\rho_0=N/V$ denotes the density of the themometer.

In conclusion, we explore correlation-enhanced low-temperature quantum thermometry leveraging a Ramsey interferometry protocol. By sparsely immersing $N$ thermometers in a common low-temperature bath, we demonstrate that Heisenberg scaling in estimation precision can be achieved, even starting from an initial product state and utilizing ensemble measurements. This observation is based on the assumption that the cooperative decay induced by the quantum noise only occurs in the low-frequency regime, a condition met by the sparse density of thermometers in cold-atom systems. This indicates that such precision enhancement is attainable in cold-atom systems.

The work is supported by the National Natural Science Foundation (Grant No. 12247101) and Fundamental Research Funds for the Central Universities (Grant No. 561219028).

\appendix 
\section{probability distribution-I} \label{app:A-I}
Consider the density matrix $\rho_{\bm{\eta},\bm{\bar{\eta}}}$  shown in the maintext Eq. \eqref{eq:rho}. The measurement along the $\pm \vec{e}_{\theta}$ on all thermometers yields 
\begin{align}
P_{\bm{s}}=&\sum_{\bm{\eta},\bm{\bar{\eta}}}\rho_{\bm{\eta},\bm{\bar{\eta}}} \prod^{N}_{j=1}\text{Tr}[\hat{\Pi}_{\theta,s_j} |\eta_{z,j}\rangle \langle \bar{\eta}_{z,j}|],
\end{align}
where $\hat{\Pi}_{\theta,s_j}=\frac{1}{2}( |+_z\rangle \langle +_z|+ |-_z\rangle \langle -_z|+s_j e^{-i\theta}|+_z\rangle \langle -_z|+s_j e^{i\theta}|-_z\rangle \langle +_z|)$ and $s_j=\pm$ denotes the readout of $j$-th thermometer. After some simplifications, one can get that
\begin{equation}
 P_{\bm{s}}= \sum_{\bm{\eta},\bm{\bar{\eta}}}\frac{1}{2^N}\rho_{\bm{\eta},\bm{\bar{\eta}}} e^{i  \sum_j (\theta \Delta \eta_j + \frac{1-s_j}{2} \pi)}, 
\end{equation}
where $s_j=e^{i \frac{1-s_j}{2} \pi}$ is used.  Using the detail expression of $\rho_{\bm{\eta},\bm{\bar{\eta}}}$, one can get that 
\begin{equation}
     P_{\bm{s}}= \sum_{\bm{\eta},\bm{\bar{\eta}}}\frac{1}{4^N} e^{-\sum_{i,j} \Delta \eta_i \Gamma_{i,j} \Delta \eta_j +i  \sum_j (\theta \Delta \eta_j + \frac{1-s_j}{2} \pi)}.
\end{equation}
By further noting that only $\{\Delta \eta_i \}$ is relevant in the exponential function, the summation over $\bm{\eta}$ and $\bm{\bar{\eta}}$ can be further represented as the average over $\{\Delta \eta_j\}$, given by
\begin{equation}
     P_{\bm{s}}= \langle e^{-\sum_{i,j} \Delta \eta_i \Gamma_{i,j} \Delta \eta_j +i  \sum_j (\theta \Delta \eta_j + \frac{1-s_j}{2} \pi)}\rangle_{\Delta \bm{\eta}},
\end{equation}
where $\Delta \eta_i=-1,0,1$ with probabilities $1/4$, $1/2$, and $1/4$, respectively. 
\section{probability distribution-II} \label{app:A-II}
 Consider the the probability distribution shown in Eq. \eqref{eq:probability}, given by
\begin{equation}\label{eq:probability-app}
     P_{\bm{s}}= \langle \prod_{j} \frac{1+s_j e^{-\Gamma_H}\cos(\theta+2\varphi_0)}{2}\rangle_{\varphi_0},
\end{equation}
where $s_j$ denotes the measurement readout of $j$-th thermometer along the measurement along $\theta$ direction and $\langle \rangle_{\varphi_0}$ indicates the average over $\varphi_0$ with distribution function $\frac{1}{\sqrt{\pi\Gamma_L}}e^{-\frac{\varphi^2_0}{\Gamma_L}}$. 
When $\theta=0$, we can get that
\begin{align}
     P_{\bm{s}}= & \langle \prod_{j} \frac{1+s_j e^{-\Gamma_H} [1-2\varphi^2_0+\mathcal{O}(\varphi^4_0) ]}{2}\rangle_{\varphi_0} \nonumber \\
     =& \prod_{j} [\frac{1+s_j e^{-\Gamma_H} (1-\Gamma_L)}{2}+\mathcal{O}(\Gamma^2_L)]
     \nonumber \\
     =&  \prod_{j} [\frac{1+s_j e^{-\Gamma}}{2}+\mathcal{O}(\Gamma^2_L)],
\end{align}
where $\Gamma=\Gamma_L+\Gamma_H$ and the relation $\langle \varphi^2_0 \rangle_{\varphi_0}=\frac{\Gamma_L}{2}$ is used.  When $\theta=\pi/2$, we can get that
\begin{align}
     P_{\bm{s}}= & \langle \prod_{j} \frac{1-s_j e^{-\Gamma_H} [2\varphi_0-\frac{4\varphi_0^3}{3}+\mathcal{O}(\varphi^5_0)]}{2}\rangle_{\varphi_0} \nonumber \\
     =& \langle \prod_{j}[\frac{1-s_j e^{-\Gamma_H} 2\varphi_0(1-\Gamma_L)}{2}+\mathcal{O}(\Gamma^3_L)]\rangle_{\varphi_0} \nonumber \\
     =&  \frac{1}{2^N}[\langle \prod_{j}[e^{-2s_j e^{-\Gamma} \varphi_0 -2 e^{-2\Gamma}\varphi^2_0}+\mathcal{O}(\Gamma^2_L)]\rangle_{\varphi_0}] \nonumber \\
     =& \frac{1}{2^N}[\frac{e^{S^2 \frac{\Gamma_L e^{-2\Gamma}}{1+2N\Gamma_L e^{-2\Gamma}}}}{\sqrt{1+2N\Gamma_L e^{-2\Gamma}}}+\mathcal{O}(N\Gamma^2_L)],
\end{align}
where $S=\sum_j s_j$ and relations $ \varphi^3_{0}=\frac{3}{2}\Gamma_L \varphi_0+\mathcal{O}(\Gamma^3_L)$ and $1-s_j e^{-\Gamma} 2\varphi_0=e^{-s_j e^{-\Gamma} 2\varphi_0-2 e^{-2\Gamma} \varphi^2_0}+\mathcal{O}(\varphi^3_0)$ are used.
\bibliography{TU}

\end{document}